# FastMRI Prostate: A Publicly Available, Biparametric MRI Dataset to Advance Machine Learning for Prostate Cancer Imaging

Radhika Tibrewala[1,2], Tarun Dutt[1,2], Angela Tong[1,2], Luke Ginocchio[2], Mahesh B Keerthivasan[2,3], Steven H Baete[1,2], Sumit Chopra[2,4], Yvonne W Lui[1,2], Daniel K Sodickson[1,2], Hersh Chandarana[1,2], Patricia M Johnson[1,2]

[1] Center for Advanced Imaging Innovation and Research (CAI$^2$R), Department of Radiology, New York University Grossman School of Medicine, New York, New York, USA.
[2] Bernard and Irene Schwartz Center for Biomedical Imaging, Department of Radiology, New York University Grossman School of Medicine, New York, New York, USA.
[3] Siemens Medical Solutions USA, New York, NY, USA
[4] Courant Institute of Mathematical Sciences, New York University, New York, USA

## Abstract

The fastMRI brain and knee dataset has enabled significant advances in exploring reconstruction methods for improving speed and image quality for Magnetic Resonance Imaging (MRI) via novel, clinically relevant reconstruction approaches. In this study, we describe the April 2023 expansion of the fastMRI dataset to include biparametric prostate MRI data acquired on a clinical population. The dataset consists of raw k-space and reconstructed images for T2-weighted and diffusion-weighted sequences along with slice-level labels that indicate the presence and grade of prostate cancer. As has been the case with fastMRI, increasing accessibility to raw prostate MRI data will further facilitate research in MR image reconstruction and evaluation with the larger goal of improving the utility of MRI for prostate cancer detection and evaluation. The dataset is available at https://fastmri.med.nyu.edu.

## 1    Introduction

Recent years have seen a surge in machine-learning-based MR image reconstruction research [1-4]. Supervised machine-learning-based methods for image reconstruction require large quantities of raw k-space data for model training. Public MRI datasets such as MRNet [5], HCP [6], ADNI [7], and OAI [8] are limited to reconstructed images and do not include raw k-space data. The limited availability of raw k-space datasets motivated the release of fastMRI (brain and knee) in 2020 and SKM-TEA (knee) in 2022 [9, 10]. In 2021, the fastMRI+ initiative released bounding box annotations and study-level labels for various pathologies in the brain and knee, enabling the development of clinically relevant reconstruction and detection models [11]. In this paper, we describe the April 2023 expansion of fastMRI to include biparametric prostate MRI data acquired in a clinical population.

Prostate cancer (PCa) is the most diagnosed malignancy and the fourth leading cause of death in men[12]. Biparametric MRI with T2-weighted and diffusion-weighted sequences is an effective diagnostic tool used in PCa management [13-15]. Faster imaging using advanced reconstruction of under-sampled data may enable more cost-effective workflows for prostate MRI, and automated triage may help to make prostate MRI more efficient and more widely accessible. To further advance these goals, we have added a prostate dataset to fastMRI to facilitate the development of machine learning tools which may increase the utility of prostate MRI. The dataset consists of T2-weighted and diffusion-weighted k-space data acquired on clinical 3T MRI systems, along with the reconstructed images and slice-level labels which indicate the presence and grade of prostate cancer.

## 2 Materials and Methods

### 2.1 Patient population

This dataset includes data from 312 male patients referred for clinical prostate MRI at NYU Langone Health (NYULH) between March 2020 and April 2021. The mean age of the subjects was $66 \pm 8$ years. The curation of the dataset was part of a study approved by our local institutional review board.

### 2.2 Description of dataset

All data were de-identified via conversion to the vendor-neutral International Society for Magnetic Resonance in Medicine raw data format (ISMRMRD). Data were acquired on two clinical 3T systems (MAGNETOM Vida, Siemens Healthineers, Germany), each with a matrix coil array made up of selected channels from an anterior body coil array and a posterior spine coil array (See Table 2 for range of coil element numbers selected). 2D axial T2-weighted turbo spin echo (TSE) and echo planar imaging (EPI)-DWI sequences were acquired for each patient. The T2-weighted data were acquired with 3 averages, the first and third averages sampling the odd lines of k-space, and the second average sampling the even lines of k-space. The EPI-DWI scans were performed using tri-directional diffusion-sensitizing gradients with $b$ values of 50 s/mm$^2$ (B50) and 1000 s/mm$^2$ (B1000), performed with 4 and 12 averages, respectively. The parallel imaging acceleration factor for EPI-DWI scans was $R = 2$. Additional scan parameters for both sequences are listed in **Table 1**.

Table 1. Scan parameters for the axial T2W TSE sequence and the EPI-DWI sequence

| Scan Parameter | T2W | DWI |
|---|---|---|
| TR (s) | 3.5-7.2 | 5.0–7.3 |
| TE (ms) | 100 | 77 |
| ETL | 25 | 75 |
| In-plane resolution (mm)* | 0.56 x 0.56 | 2.0 x 2.0 |
| Slice thickness (mm) | 3 | 3 |
| Matrix size | 320 x 320 | 100 x 100 |
| FOV (mm) | 180 x 180 | 200 x 200 |
| Number of slices | 30-36 | 24-38 |
| Number of receive channels** | 10-30 | 14-30 |

\* these acquisitions have a phase resolution of 0.7, so the acquired data has a resolution of 0.56 x 0.80 mm which is then interpolated to 0.56 x 0.56 mm
\*\* combination of a body coil and a spine coil is used in acquiring this data, however, for a given scan, all coil elements may not be used depending on patient size and position, creating a variation in the number of receive channels

The 312 patient acquisitions were divided into training/validation/test groups (218/48/46 patients). The data included in each patient file is described in **Table 2**.

**Table 2**. Description of data fields of fastMRI prostate files

| Sequence | Field | Description |
| --- | --- | --- |
| T2 | ismrmrd_header | Anonymized metadata in the ISMRMRD standard data format |
| | kspace | Raw k-space data with dimensions (averages, slices, coils, readout, phase) |
| | reconstruction_rss | Reconstructed image volume with dimensions (slices, x, y) |
| | calibration_data | Calibration scan acquired with dimensions (slices, coils, readout, phase) |
| DWI | ismrmrd_header | Anonymized metadata in the ISMRMRD standard data formats |
| | kspace | Raw k-space data with dimensions (diffusion direction, slices, coils, readout, phase) |
| | b50x, b50y, b50z | Reconstructed source image with $b = 50$ s/mm$^2$ in the X, Y and Z directions with dimensions (slices, x, y) |
| | b1000x, b1000y, b1000z | Reconstructed source image with $b = 1000$ s/mm$^2$ in the X, Y and Z directions with dimensions (slices, x, y) |
| | trace_b50, trace_b1000 | Geometric mean of tri-directional source images with dimensions (slices, x, y) |
| | adc_map | Apparent diffusion coefficient map, estimated from the source images with dimensions (slices, x, y) |
| | calibration_data | Calibration scan acquired with dimensions (slices, coils, readout, phase) |
| | coil_sens_maps | Coil sensitivity maps calculated using ESPIRIT with dimensions (slices, coils, readout, phase) |
| | phase_correction | Phase correction scan acquired with dimensions (diffusion direction, slices, coils, readout, polarity*) |
| | b1500 | Calculated source image with b = 1500s/mm2 with dimensions (slices, x, y) |

*\* bipolar EPI readout trajectory*

## 2.3 Image reconstruction

Reconstructed images are included in the provided data files. Python scripts to reconstruct the T2 and DWI images from the k-space data are included in the public GitHub repository: https://github.com/cai2r/fastMRI_prostate. For the T2-weighted data, each under-sampled average is reconstructed using GeneRalized Autocalibrating Partially Parallel Acquisition (GRAPPA) [16] with a root sum of squares coil combination. The three averages are then combined in image space. The DWI reconstruction includes EPI gridding, GRAPPA reconstruction, and an SNR-optimizing matched filter coil combination [17] with coil sensitivities estimated using ESPIRIT [18]. The trace, ADC, and estimated B1500 are computed from the six individual source images.

**Figure 1** shows a single slice of the T2-weighted image set, along with a B50 trace, B1000 trace, and ADC map for each of two subjects with PI-RADS scores 1 and 5, respectively.

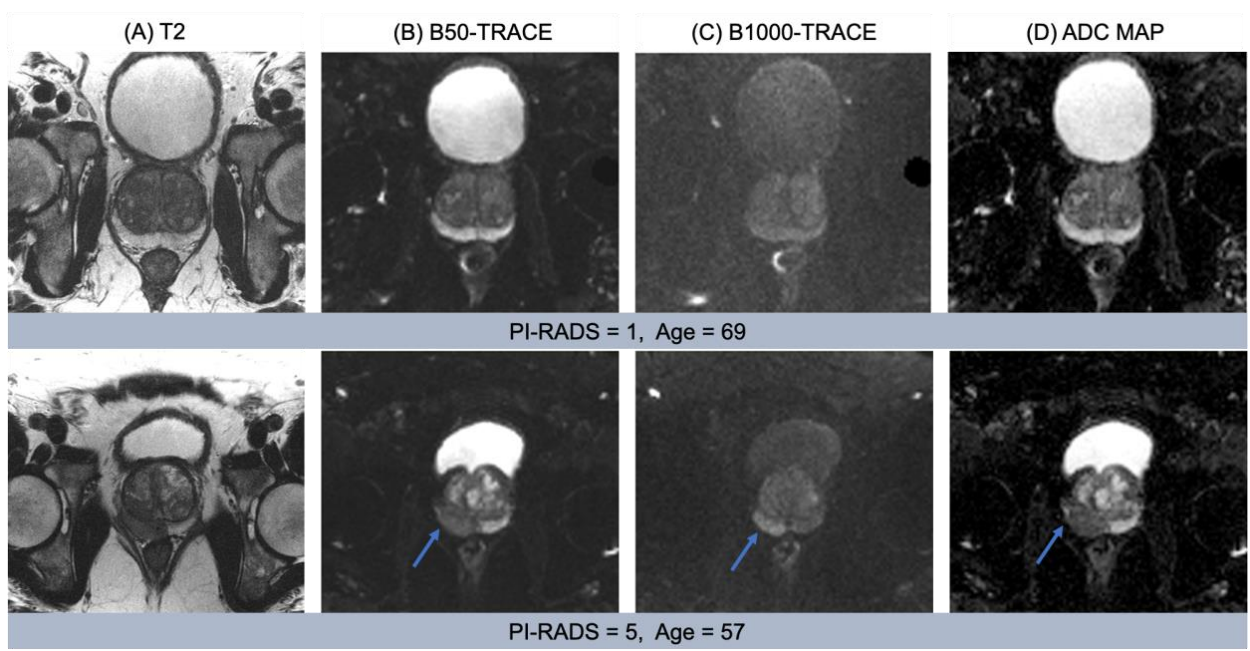

*Figure 1: (Row 1) A reconstructed axial T2-image (a), B50 DWI (b), B1000 DWI (c) and ADC Map (d) for a 69-year-old man referred for prostate evaluation, with a PI-RADS assessment of 1 based on the full set of acquired MR images. (Row 2) A reconstructed axial T2-image (a), B50 DWI (b), B1000 DWI (c) and ADC Map (d) for a 57-year-old man referred for clinical prostate MRI, with a PI-RADS assessment of 5. Blue arrows indicate a lesion in the right posteromedial midgland peripheral zone.*

## 2.4 Description of labels

A fellowship-trained radiologist reviewed all the T2 and DWI images, assigning a Prostate Imaging Reporting & Data System (PI-RADS v2.1) label to each slice of a patient volume on each sequence [19]. PI-RADS v2.1 assessment uses a 5-point scale based on the likelihood that findings correlate with the presence of a clinically significant cancer with the following categories: PIRADS 1 - Very Low (highly unlikely it is clinically significant cancer), PIRADS 2 - Low (unlikely it is clinically significant cancer), PIRADS 3 - Intermediate (uncertain if it is clinically significant cancer), PIRADS 4 - High (likely that it is clinically significant cancer), PIRADS 5 - Very High (highly likely that it is clinically significant cancer). The images were labeled retrospectively for research purposes; the labels were not utilized to direct clinical care.

## 3 Discussion

To our knowledge, this is the only available raw k-space dataset that includes a diffusion MR sequence from a clinical population. As has been the case for fastMRI+, the addition of labels will facilitate task-based assessment of image reconstructions, development of end-to-end methods (e.g., moving directly from k-space to lesion detection), and/or joint training of reconstruction and classification models, with the larger goal of improving the utility of MRI for prostate cancer detection and evaluation.

In some cases, our offline reconstruction (in the provide Python script) has inferior image quality compared to the inline reconstruction which uses proprietary algorithms on commercial scanners. We encourage the community to explore improvements to this offline reconstruction in terms of both image quality and computational efficiency.

The acquisition strategy for T2-weighted data, with three under-sampled averages, is slightly unconventional, but was selected for historical reasons in our institution's clinical protocol. Rather than discarding averages or pre-combining them, we have chosen to provide all three averages for maximal flexibility in using this data.

Another noteworthy feature of this data release involves the PI-RADS labeling. The T2 and DWI images were both labeled slice-by-slice in isolation. This is a deviation from how clinical interpretation of these images is performed, where a PI-RADS score would be assigned to each subject using information from the entire exam. For this dataset, we chose to provide slice-level labels because this results in a training dataset size on the order of the number of slices rather than the number of exams (with the caveat that nearby slices in the same patient will be correlated). Volume-level labels for each sequence can be derived using a max operation between slice-level scores. Exam level labels (in accordance with clinical practice) can be derived by performing a max operation between sequence scores. As we expand this prostate dataset, we expect that the power of volume-level labels more closely matching typical clinical assessments will increase. In the meantime, slice-level labels can still provide meaningful information about the conspicuity of a lesion in the volume in which it was labeled.

Finally, this prostate dataset has fewer subjects (312) compared to the brain (5404) and knee (1290) fastMRI datasets, but it includes a diffusion sequence, which adds value to the existing fastMRI collection. We expect to expand the number of prostate imaging subjects in the future. In the meantime, we hope that the provision of this dataset will further facilitate research in MR image reconstruction and evaluation and will serve as a benchmark for training new machine learning tools devoted to prostate cancer imaging.

# 4    Acknowledgments

The work reported in this publication was supported by the Center for Advanced Imaging Innovation and Research (CAI$^2$R), a National Center for Biomedical Imaging and Bioengineering operated by NYU Langone Health and funded by the National Institute of Biomedical Imaging and Bioengineering through award number P41EB017183 and R01EB024532.